\documentclass[prd,aps,showpacs,byrevtex]{revtex4}

\usepackage{graphicx}
\usepackage{bm}

\def\ov{\overline}
\def\be{\begin{equation}}
\def\ee{\end{equation}}
\def\ba{\begin{eqnarray}}
\def\ea{\end{eqnarray}}

\def\op{\oplus}
\def\al{\alpha}
\def\Z{M_Z}

\def\R{M_R}
\def\U{M_U}
\def\D{\Delta}
\def\T{\Theta}
\def\ep{\epsilon}
\def\br{\begin{array}}
\def\er{\end{array}}
\def\bc{\begin{center}}
\def\ec{\end{center}}
\def\gsim{\mathop{\smash{>}}\limits_\sim}

\def\lr{{\cal G}_{LR}}
\def\std{{\cal G}_{std}}

\def\bpmat{\begin{pmatrix}}
\def\epmat{\end{pmatrix}}

\begin{document}

\setcounter{footnote}{0}
\renewcommand{\thefootnote}{\fnsymbol{footnote}}
 
\title{ Low Intermediate Scales for Leptogenesis in Supersymmetric\\
 ${\bf SO(10)}$ Grand Unified Theories}

\author{Swarup Kumar Majee $^{a,b,}$}
\email{swarup@mri.ernet.in}
\author{Mina K. Parida$^{a,}$}
\email{paridam@mri.ernet.in}
\author{Amitava Raychaudhuri${}^{a,b}$}
\author{Utpal Sarkar $^{c,}$}
\email{utpal@prl.res.in}
\vspace{10pt}

\affiliation{ ${}^{a)}$ Harish-Chandra Research Institute, Jhunsi, Allahabad
211 019, India \\
   ${}^{b)}$ Department of Physics, University of Calcutta, 
92 A.P.C. Road, Kolkata 700 009, India\\ 
   ${}^{c)}$ Physical Research Laboratory, Ahmedabad 380 009, India}


\begin{abstract}


A low intermediate scale within minimal supersymmetric $SO(10)$
GUTs is a desirable feature to accommodate leptogenesis. We
explore this possibility in models where the intermediate gauge symmetry 
breaks spontaneously by (a) doublet Higgs scalars and also (b) by triplets.
 In both scenarios gauge coupling 
unification requires the scale of left-right symmetry breaking ($M_R$) to
be close to the unification scale.  This will entail unnaturally
small neutrino Yukawa couplings to avoid the gravitino problem
and allow successful leptogenesis.  We point out that any one of
three options --  threshold corrections due to the mass spectrum
near the unification scale, gravity induced non-renormalizable
operators near the Planck scale, or presence of additional light
Higgs multiplets -- can permit unification along with 
much lower values of $M_R$ as required for leptogenesis. In the 
triplet model, independent of these corrections, we find a lower bound 
on the intermediate scale,
 $M_R > 10^9$ GeV, arising from the requirement that the theory
must remain perturbative at least upto the GUT scale. We show that
in the doublet model $M_R$ can even be in the TeV region which,
apart from permitting resonant leptogenesis, can be tested at LHC and ILC.

\end{abstract}

\pacs{12.10.Dm, 11.10.Hi }
\maketitle

\renewcommand{\thesection}{\Roman{section}}
\setcounter{footnote}{0}
\renewcommand{\thefootnote}{\arabic{footnote}}

\section{Introduction}

An area where the standard model based on the group $SU(3)_C
\times SU(2)_L \times U(1)_{Y} \equiv {\cal G}_{std}$ merits
improvement is the origin of parity violation. The most natural
extension that addresses this issue is the left-right symmetric
model in which the gauge group is enlarged to $SU(3)_C \times
SU(2)_L \times SU(2)_R \times U(1)_{(B-L)} \equiv {\cal G}_{LR}$
\cite{lr}. Here, the left-handed fermions transform nontrivially
under $SU(2)_L$ and are singlet under $SU(2)_R$, while it is the
converse for the right-handed fermions.  It is then possible to
extend the definition of parity of the Lorentz group to all
particles and ensure that the theory is invariant under the
transformation of parity. Spontaneous breaking of the group
$SU(2)_R$ would trigger violation of parity in the low energy
theory. It is also possible to break the parity symmetry
spontaneously by the vacuum expectation value ($vev$) of a gauge
singlet scalar field which has odd parity \cite{dpar}.  In either
case, parity violation at low energy originates from some
spontaneous symmetry breaking at high energy.

The simplest grand unified theory (GUT) that  includes the
left-right symmetric extension of the standard model is based on
the gauge group $SO(10)$ and has been studied very widely
\cite{so10}.  In recent times there is renewed interest in the
$SO(10)$ GUT stemming from the predictability of the minimal
structure of the models \cite{min1}. These minimal $SO(10)$
models with the most economical choices of  Higgs scalars have
several interesting features
\cite{min1,min2,min3,min4}. Here we show that the possibility of
leptogenesis ~\cite{lept,lepto} can also be accommodated in these
models.  From an analysis of gauge coupling unification, we
determine the scale of left-right symmetry breaking, which is
intimately related to a successful prediction of leptogenesis in
these models. An apparent obstacle arises in the following form:
either these models do not allow any intermediate mass scales or
the intermediate left-right symmetry breaking scale comes out to
be large ($\sim 10^{15}$ GeV).  To implement leptogenesis,  on
the other hand, the left-right symmetry breaking scale has  to be
much lower. We exhibit several alternate possibilities which may
provide a way out from this impasse.

There are two broad classes of minimal $SO(10)$ models: those
with only doublet Higgs scalars (Model I) and the conventional
left-right symmetric model including triplet Higgs scalars (Model
II).  In both versions, a bi-doublet Higgs scalar ($\phi \equiv (1,2,2,0)$
under ${\cal G}_{LR}$) 
 gives mass to the charged fermions and
also a Dirac mass to the neutrinos \cite{f1}.
In an $SO(10)$ GUT, this bi-doublet $\phi$ belongs to the
representation ${\bf 10, 120}$ or ${\bf 126}$. Usually a ${\bf
10}$ representation is chosen. However, for correct
fermion mass relations \cite{gj}, a ${\bf \overline {126}}$ representation
containing the field $\phi' \equiv \{15,2,2\}$ under the group
$SU(4)_C \times SU(2)_L \times SU(2)_R \equiv {\cal G}_{PS}$
is often chosen. 

The main differences between Models I and II lie in the Higgs
scalar that breaks the left-right symmetry and the generation of
neutrino masses. Lepton number violation in these models arises
from the Higgs scalars that break the $B-L$ symmetry and hence
the left-right symmetry. The origin of leptogenesis is also
different in these two models.  There is a natural mechanism of
resonant leptogenesis in Model I (see below) while 
Model II has other advantages.

In Model I, the left-right symmetric group ${\cal G}_{LR}$ is
broken by an $SU(2)_R$ doublet Higgs scalar $\chi_R
\equiv (1,1,2,-1)$ when its neutral component acquires a $vev$
$\langle \chi^\circ_R \rangle \sim v_R$.  Left-right parity
implies the presence of an $SU(2)_L$ doublet Higgs scalar $\chi_L
\equiv (1,2,1,-1)$. The $vev$ of the neutral component of this
field,  $\langle \chi^\circ_L \rangle \sim v_L$, in addition to
$\langle \phi \rangle$, breaks the
electroweak symmetry.

In this Model there is an extra singlet fermion, $S$, that combines
with the neutrinos and a new type of see-saw mechanism is
operational \cite{valle}. There are several interesting features
associated with this. The one relevant here is that the singlet
fermions can be almost degenerate with the neutrinos, leading to
resonant leptogenesis naturally in this scenario \cite{new}.

In Model II, an $SU(2)_R$ triplet Higgs scalar $\bar {\Delta}_R
\equiv (1,1,3, 2)$ breaks the left-right symmetric group ${\cal
G}_{LR}$. When the neutral component acquires a $vev$,
$\langle{\bar{\Delta}}^\circ_R \rangle \sim v_R$, it gives
Majorana masses to the right-handed neutrinos breaking lepton
number by two units. When the bi-doublet Higgs scalar $\phi$
breaks the electroweak symmetry, this leads to the small see-saw
neutrino mass \cite{see-saw}.  Due to left-right parity, there is also an
$SU(2)_L$ triplet Higgs scalar $\bar {\Delta}_L \equiv (1,3,1,
2)$. Although these scalars have a mass at the parity
breaking scale $M_R$, the $vev$ of the neutral component of this
field is extremely tiny and can give small Majorana  masses to
the left-handed neutrinos leading to the type II see-saw
mechanism, explanation of large neutrino mixings through $b-\tau$
unification, and  parameterization of all fermion masses,
mixings, and CP-violation.  Decays of the right-handed neutrinos
or the left-handed triplet Higgs scalars can generate a lepton
asymmetry at the scale $M_R$. With high left-right symmetry
breaking scale and asymptotic parity conservation, Model II is
truly a renormalizable high scale SUSY $SO(10)$ theory of fermion
masses and mixings \cite{min1, min2, min3, goh}.

The Majorana mass of right-handed neutrinos is given by $M_N \sim
\tilde{f} v_R$, where $\tilde{f}$ is the Yukawa coupling.  The
right-handed neutrino mass-scale controls leptogenesis as well as
light neutrino masses and, in particular, a value around $10^{9}$
GeV or lower is favored by the `gravitino constraint' discussed
below.  Since $\tilde{f}$ does not affect the experimentally
measured charged fermion masses at low energies, one can assign
any value to it, leaving the left-right symmetry breaking scale
unrestricted. However, such a low RH neutrino mass is
likely to give too large contributions to the left-handed neutrino
masses through the see-saw mechanism,  contradicting experimental
observation. The main motivation of the see-saw mechanism was to
avoid arbitrarily small Yukawa couplings, so we shall assume the
value of $\tilde{f}$ to be of order unity \cite{f2}.

While considering leptogenesis in the minimal supersymmetric
$SO(10)$ GUTs, the potential problem \cite{grav} arising from the
overclosure of the universe by gravitinos (and its adverse influence on
the successful Big Bang Nucleosynthesis predictions) must be taken into
account.  This requires the reheating temperature, $T_{RH}$, to
be less than $\sim 10^8$~GeV.  Since leptogenesis takes place
just below the scale of left-right symmetry breaking, $M_R >
T_{RH}$ can make models inconsistent with the above or at least
unnatural.  However, Model I may still be consistent because it
offers the alternative of resonant leptogenesis.\\
\par

Using renormalization
group (RG) equations, in the following sections we examine for
both Models whether gauge coupling unification at all allows a
low left-right symmetry breaking scale which would make
successful leptogenesis viable.


Models I and II have the same symmetry breaking chain:
\begin{eqnarray}
SO(10) & \stackrel{210 ~(M_U)}{\longrightarrow} &  SU(3)_C \times
SU(2)_L \times SU(2)_R \times U(1)_{B-L}  \nonumber \\
&\stackrel{16 ~{\rm or} ~126 ~(M_R)}{\longrightarrow}&SU(3)_C
\times SU(2)_L \times U(1)_Y  \nonumber \\ &\stackrel{10
~(M_Z)}{\longrightarrow}&SU(3)_C \times U(1)_Q \nonumber
\end{eqnarray}

At the GUT scale, the symmetry is broken by the vacuum
expectation value  of a ${\bf 210}$ dimensional representation of
$SO(10)$.  The ${\bf 210}$ has a singlet under the  subgroup ${\cal
G}_{PS}$, i.e., $\{1,1,1\}$, which is odd under parity. When this
field acquires a $vev$, $SO(10)$ is broken to  ${\cal G}_{PS}$
and D-parity is also spontaneously broken (i.e., $g_{2L} \neq g_{2R}$).
To keep D-parity intact at this level we have to look elsewhere.
The $SO(10)$ {\bf 210} also contains a \{15,1,1\} under  ${\cal
G}_{PS}$ which is D-parity even. This is the field to which the
$vev$ must be ascribed to get the desired symmetry breaking to
${\cal G}_{LR}$ while keeping D-parity intact.

The left-right symmetry, $\lr$, is broken by the $vev$ of the
fields $F + \bar F$, where $F$ is a ${\bf {16}(\equiv\bf {\Gamma})}$ dimensional
representation for Model I and a ${\bf 126}$ dimensional
representation for Model II. Finally, the electroweak symmetry
breaking takes place by the $vev$ of a {\bf 10}-plet of $SO(10)$.
In the minimal models under consideration, there are no other
Higgs representations.

The simplicity of the minimal supersymmetric $SO(10)$ GUT allows
several interesting predictions. With some standard assumptions
it is possible to determine the mass scales involved in the
symmetry breaking.  Below, we shall  show that one-loop
renormalization group evolution leads to left-right
symmetry breaking and unification scales,
\ba &&\R^0 \simeq 1.3\times 10^{16}~ {\rm GeV},~~~~\U^0 \simeq 2.9\times
10^{16}~ {\rm GeV} .
\ea
$\R^0$ and $\U^0$ are already very close. But, the situation
worsens when two-loop RG contributions are included and we find
that no intermediate scales are allowed at all below the
unification scale.  All this makes leptogenesis unnatural in this
class of models. We suggest some possible remedies.


\par
In this paper we show that inclusion of GUT-threshold effects, gravitational
corrections through $\rm {dim}.5$ operators, or presence of additional light
fields near  $M_R$, can lower the intermediate scale, bringing it even
to the range of a few TeV in the doublet model. Thus, in this
model, the gravitino constraint can be easily satisfied
leading to successful resonant leptogenesis at low scales. In
addition, the signatures of right handed gauge bosons,
($W_R^{\pm}$, $Z_R$), and new Higgs scalars can be tested at the LHC
and ILC.  In the triplet model, on the other hand, even though the GUT
threshold corrections are much larger, we derive a 
bound on the intermediate scale, $M_R > 10^{9}$ GeV arising out
of the requirement of perturbation theory to be valid, due to 
which the scale cannot be reduced further. With this lower bound
on $M_R$ the triplet model emerges genuinely as a high scale
supersymmetric theory for successful description of fermion
masses and mixings.

\par
This paper is organised in the following manner. In Sec.II we discuss
renormalization group equations and origins of threshold and
Planck scale effects.
In Sec.III we show how low intermediate scales are obtained in the doublet
model and $M_R$ can be lowered upto $5\times 10^{9}$ GeV in the
triplet model. The perturbative lower bound on $M_R$ is derived
in Sec.IV. After making  brief remarks on fermion masses and light
scalars in the SUSY $SO(10)$ model in Sec.V, we summarise the
results and state our conclusions in Sec.VI.
  
\section{Renormalization group equations}

In this section, first we present the RG equations including (a) 
two-loop beta functions, (b) threshold effects, and (c) 
contributions from non-renormalizable interactions appearing at
the Planck scale. Later we show that, with the minimal particle
content and  in the absence of effects due to (b) and (c), at the
one-loop level Models I and II imply a high scale for $M_R$ near
$M_U$ but when two-loop effects are taken into account even that
is excluded.

\subsection{General formulation}

The RG equations with one \cite{f3}  intermediate scale, $M_R$, between $M_U$ and $M_Z$ are:
\ba
{1\over\al_i(\Z)}&=&{1\over\al_i(\R)}+{a_i\over
2\pi}\ln{\R\over\Z}+\T_i-\D_i,
\label{eqA.1}\\
{1\over\al_i(\R)}&=&{1\over\al_i(\U)}+{a'_i\over
2\pi}\ln{\U\over\R}+\T'_i-\D'_i-
\D_i^{(gr)}, \nonumber \\
\label{eqA.2}
\ea
\noindent
where $i$ runs over the different gauge couplings. In the R.H.S.
of eqs. (\ref{eqA.1}) and (\ref{eqA.2}), the second and third
terms represent one- and two-loop contributions, respectively,
with
\ba
\T_i &=&{1\over 4\pi}\sum_jB_{ij}\ln{\al_j(\R)\over\al_j(\Z)},\nonumber\\
\T'_i&=&{1\over 4\pi}\sum_jB'_{ij}\ln{\al_j(\U)\over\al_j(\R)},\nonumber\\
B_{ij}&=&{b_{ij}\over a_j},\, B'_{ij}={b'_{ij}\over a'_j}.\label{eqA.3}
\ea
\par\noindent
The one- and two-loop coefficients ($a_j, a'_j, b_{ij}, b'_{ij}$)
for specific scenarios are given later.   Between $M_Z$ and $M_R$
the indices $i,j \subset \std $ while above $M_R$ one has  $i,j
\subset \lr$.

 The $\D_i$ include SUSY threshold effects  and
intermediate scale threshold effects at $\R$,
$$\D_i=\D^{(S)}_i+\D^{(R)}_i,$$ while $\D'_i$ includes the same at the unification scale $\U$.
They are represented as \cite{carena, baer, langacker, parida1},
\ba
\D_i^{(S)} &=& \frac{1}{2\pi}\Sigma_{\al} b_i^\al
\ln{{M^{\al}}\over M_S} \equiv
\frac{b_i} {2\pi}\ln{{M_i}\over {M_S}}, \;\;\;
b_i =  \Sigma_{\al} b_i^\al,   \nonumber\\
\D_i^{(R)} &=& \frac{1}{2\pi}\Sigma_{\beta}  {c}_i^\beta
\ln{{M^{\beta}}\over M_R} \equiv
\frac{b'_i} {2\pi}\ln{{M_i}\over {M_R}}, \;\;\;
b'_i = \Sigma_{\beta}  {c}_i^\beta , \nonumber\\
\D'_i &=& \frac{1}{2\pi}\Sigma_{\gamma} {d}_i^\gamma
\ln{{M^{\gamma}}\over M_U} \equiv
\frac{b''_i} {2\pi}\ln{{M_i}\over {M_U}}, \;\;\;
b''_i = \Sigma_{\gamma} {d}_i^\gamma . ~~~~ \label{eqA.4} \\
\D_i^{(gr)} &=& -{{\epsilon}_i\over {\al}_G}, i = BL, 2L, 2R ,3C
.\label{eqA.4a}
\ea
Here the indices $\al,\beta$ and $\gamma$ signify the particle
components of $SO(10)$ representations spread around the  SUSY
scale  $M_S$, the $SU(2)_R\times U(1)_{B-L}$ breaking scale
$M_R$, and the $SO(10)$ breaking scale $M_U$, respectively.

 The definition of effective mass parameters at the SUSY
scale $M_S$ through the first of eqs. (\ref{eqA.4})  introduced by
Carena, Pokorski and Wagner~\cite{carena} has been generalised to
study GUT-threshold effects by Langacker and Polonsky
\cite{langacker} in SUSY $SU(5)$ and in ref.~\cite{parida1} to study
 intermediate breaking in SUSY $SO(10)$. The effective mass
parameters defined through these relations  are not arbitrary.
Logarithm of each of them is a well defined
linear combination of logarithms of actual  particle masses
(heavy or superheavy) spread around the respective thresholds.
Hence, in
principle, it is possible to express them
in terms of the parameters of the superpotential.~The actual
relationship would vary from model to model 
depending upon the type and number of representations used in
driving the spontaneous symmetry  breaking of SUSY $SO(10)$ to
the low energy theory.

In the absence of unnatural mass spectra,  the particles are
expected  to be a few times heavier or lighter than the
associated threshold scale which would  result in the effective
mass parameters bearing a similar relationship to that scale.\\

The term
$\D^{gr}_i$ represents the effect of $\rm {dim}.5$-operators
which may be induced at the Planck scale. These operators
modify the boundary condition at $\U$ as \cite{shafi, parpat},
\ba
&&\al_{2L}(\U)(1+\ep_{2L})=\al_{2R}(\U)(1+\ep_{2R})
=\al_{BL}(\U)(1+\ep_{BL})=\al_{3C}(\U)(1+\ep_{3C})=\al_G.~~~~~\label{eq13a}
\ea 
Here, $\al_G = g^2(M_U)/{4\pi}$ is the GUT fine-structure
constant. The impact of various contributions in eqs.
(\ref{eqA.4}) and (\ref{eqA.4a}) in lowering the intermediate
scale in SUSY $SO(10)$ GUTs will be discussed in detail in
subsequent sections.

Using eqs. (\ref{eqA.1}) -  (\ref{eqA.4a})
one obtains for the mass scales
\cite{parida1},
\ba
\ln{\R\over \Z}&=&{1\over(AB'-A'B)}[(AL_S-A'L_\T)+(A'J_2-AK_2) 
-{2\pi\over\al_G}(A\ep''-A'\ep')+(A'J_\D-AK_\D)],\label{eqA.5}\\
\ln{\U\over\Z}&=&{1\over (AB'-A'B)}[(B'L_\T-BL_S)+(BK_2-B'J_2)
-{2\pi\over\al_G}(B'\ep'-B\ep'')+(BK_\D-B'J_\D)],\label{eqA.6}
\ea

 where
\ba L_S&=&{2\pi\over\al(\Z)}\left(1-{8\over 3}{\al(\Z)\over\al_S(\Z)}\right),
\nonumber\\
L_\T&=&{2\pi\over \al(\Z)}\left(1-{8\over 3}\sin^2\T_W(\Z)\right),\nonumber\\
A&=&a'_{2R}+{2\over 3}a'_{BL}-{5\over 3}a'_{2L},\nonumber\\
B&=&{5\over 3}(a_Y-a_{2L})-A ,
\nonumber\\
A'&=&\left(a'_{2R}+{2\over 3}a'_{BL}+a'_{2L}-{8\over
3}a'_{3C}\right),\nonumber\\
B'&=&{5\over 3}a_Y+a_{2L}-{8\over 3}a_{3C} - A' .\label{eqA.7}
\ea

\ba
J_2&=&2\pi\left[\T'_{2R}+{2\over 3}\T'_{BL}-{5\over 3}\T'_{2L}
+{5 \over 3}(\T_{Y}-\T_{2L})\right],\nonumber\\
K_2&=&2\pi\left[\T'_{2R}+{2\over 3}\T'_{BL}+\T'_{2L}-{8\over 3}\T'_{3C}\right.
+\left.{5\over 3}\T_{Y}+\T_{2L}-{8\over 3}\T_{3C}\right],\nonumber\\
\ep'&=&\ep_{2R}+{2\over 3}\ep_{BL}-{5\over 3}\ep_{2L},\nonumber\\
\ep''&=&\ep_{2L}+\ep_{2R}+{2\over 3}\ep_{BL}-{8\over 3}\ep_{3C},\nonumber\\
J_\D&=&-2\pi\left[\D'_{2R}+{2\over 3}\D'_{BL}-{5\over 3}\D'_{2L}
+{5\over 3}(\D_{Y}-\D_{2L})\right],\nonumber\\
K_\D&=&-2\pi\left[\D'_{2R}+{2\over 3}\D'_{BL}+\D'_{2L}-{8\over 3}\D'_{3C}
\right.
\left.+{5\over 3}\D_{Y}+\D_{2L}-{8\over 3}\D_{3C}\right].\label{eqA.8}
\ea

\subsection{The minimal SUSY $SO(10)$ models}

In this subsection we apply the RG
evolution detailed above to the specific minimal $SO(10)$ models
keeping only the one- and two-loop contributions in eqs.
 (\ref{eqA.1}) -  (\ref{eqA.8}).

The symmetry breaking proceeds through three steps. (a) In the
first step, the $SO(10)$ symmetry is broken at $\U$ by the $vev$ of a ${\bf
210}$ multiplet. As noted earlier, it is chosen to be along the
neutral component of $\{15,1,1\}$ under  ${\cal G}_{PS}$ which is
even under D-parity \cite{dpar}.  Thus, the gauge symmetry is
broken to ${\cal G}_{LR}$ and,  with unbroken D-parity,
left-right discrete symmetry survives preserving $g_{2L} =
g_{2R}$. (b) In the second step, in Model I (the doublet model), the
$vev$ of the neutral component of ${\ov {\chi_R}}
\subset {\bf \ov {16}}$ which transforms as $(1,1, 2,-1)$ under ${\cal
G}_{LR}$ breaks $SU(2)_R\times U(1)_{B-L} \to U(1)_Y$
at $\R$.  The left-handed doublets $\chi_L(1,2,1,-1)\op
{\ov {\chi_L}}(1,2,1,1)$ and other components of $ \chi_R (1,1,2,
-1)\op {\ov  {\chi_R}}(1,1, 2,1)$ not absorbed by the RH gauge
bosons remain light with masses around the intermediate scale
$M_R$. (c) Finally, the standard doublet Higgs contained in the
bi-doublet $\phi(1,2,2,0) \subset {\bf 10}$ drives the symmetry
breaking of ${\cal G}_{std} \to SU(3)_C \times U(1)_{em}$ at the
electroweak scale. For simplicity, in the remainder of this
section it is assumed that the supersymmetry scale, $M_S$, is the
same as $M_Z$. In Model II (the triplet model) steps (a) and (c)
remain identical. In step (b), i.e., the  breaking of  ${\cal
G}_{LR}$, the $vev$ is assigned to the neutral component of a
field $\ov{\Delta}_R \equiv$ (1,1,3,2) contained in a ${\bf
\ov{126}}$. In this alternative, the left-handed triplets
$\Delta_L(1,3,1,-2)\op {\ov {\Delta_L}}(1,3,1,2)$ contained in
the ${\bf 126}$ and ${\bf \ov{126}}$ as well as other components
of $ \Delta_R (1,1,3, -2)\op {\ov  {\Delta_R}}(1,1, 3,2)$ not
absorbed by the RH gauge bosons remain light and contribute to
the gauge coupling evolution from $M_R$.

 One major difficulty in obtaining the parity conserving ${\cal G}_{LR}$
intermediate symmetry  originates from the mass 
spectra predictions in the triplet model
with  certain colored Higgs components of $G_{PS}$ multiplets in 
$\{15,3,1\}+\{15,1,3\} \subset {\bf 210}$ being at the $M_R$
scale~\cite{min3}. 
We note that a similar  difficulty also arises in the minimal doublet model 
unless these states are made superheavy through the presence of additional 
$SO(10)$ Higgs
representations or non-renormalizable terms in the superpotential
as discussed in Sec.IV. Assuming that these additional scalars are made
superheavy, our RG analysis
applies   with the minimal particle content between $M_Z$ to $M_U$ 
as described above.

For Model I, the MSSM one- and  two-loop beta-function
coefficients below the scale $(\R)$ are given by,
\ba
&&\left(\br{c} a_{Y}\\ a_{2L}\\ a_{3C}\er\right)=\left(\br{c} {33\over 5}\\
1\\ -3\er\right), ~~~~
b_{ij}=\left(\br{ccc} {199\over 25}&{27\over 5}&{88\over 5}\\ {9\over 5}&25&
24\\
{11\over 5}&9&14\er\right),i,j \subset \std. \label{eq10}  
\ea
Above $M_R$ till $M_U$ the beta-function coefficients are
\ba
&&\left(\br{c} a'_{BL}\\ a'_{2L}\\ a'_{2R}\\ a'_{3C}\er\right)
=\left(\br{c} 9\\ 2\\ 2\\ -3\er\right),~~
b'_{ij}=\left(\br{cccc} {23/2}&{27/2}&{27/2}&{8}\\{9/2}&{32}&{3}&
{24}\\{9/2}&{3}&{32}&{24}\\{1}&{9}&{9}&{14}\er\right), i, j \subset \lr. 
 \label{eq11}
\ea
Using  $\al_S(\Z) = 0.1187$, $\al (\Z) = 1/127.9$, and
$\sin^2\T_W = 0.2312$, the one-loop solutions
yield
\ba
\R^0 = 1.3\times 10^{16}~ {\rm GeV~~~~~,} ~~~~~
\U^0 = 2.9\times 10^{16} ~{\rm GeV}.~~~~~ \label{eq12}
\ea
The GUT fine structure constant is $\al_G \simeq 1/24.25$. When
two-loop contributions are included then, as noted earlier, no
intermediate symmetry breaking scale is permitted at all. 

For Model II, below $M_R$ the one- and two-loop beta function
coefficients are still given by eq.  (\ref{eq10})
while between $M_R$ and $M_U$ we have  
\ba
&&\left(\br{c} a'_{BL}\\ a'_{2L}\\ a'_{2R}\\ a'_{3C}\er\right)
=\left(\br{c} 24\\ 5\\ 5\\ -3\er\right),~~
b'_{ij}=\left(\br{cccc} {115}&{81}&{81}&{8}\\{27}&{73}&{3}&
{24}\\{27}&{3}&{73}&{24}\\{1}&{9}&{9}&{14}\er\right), i, j \subset \lr . \label{eq11a}\ea
In this case, the one-loop evolution results in \cite{f4}
\ba
\R^0 = 7.9\times 10^{15}~ {\rm GeV~,} ~~~~~
\U^0 = 1.9\times 10^{16} ~{\rm GeV}, \label{eq12a}
\ea
with the GUT fine structure constant $\al_G \simeq 1/24.00$.
As in Model I, inclusion of two-loop effects disallows any
intermediate scale.

We shall now turn to the implication of this high intermediate
left-right symmetry breaking in the context of neutrino masses
and leptogenesis. Then we will exhibit ways by which the 
difficulties can be evaded.

\section{Low scale left-right symmetry breaking}

As noted in the previous section, in the minimal supersymmetric
$SO(10)$ models the left-right symmetry breaking intermediate scale
cannot be lower than $10^{15}$ GeV. We shall briefly illustrate the
 application of Model II for successful explanation of fermion masses
 and mixings with such a high value of $M_R$.

In Model II, the left-right symmetry is broken by the
$vev$ of the right-handed triplet Higgs scalar $\bar {\Delta}_R \equiv
(1,1,3, 2) \subset {\bf \overline {126}}$. The left-handed  
triplet Higgs scalar $\bar {\Delta}_L \equiv
(1,3,1, 2)$ required by left-right symmetry is also present in
${\bf \overline {126}}$. The bi-doublet Higgs that breaks the
electroweak symmetry and the Higgs that breaks the $SO(10)$ group
are $\phi \equiv (1,2,2,0) \subset {\bf 10}$ and $\Phi \equiv (1,1,1,0)
\subset {\bf 210}$.
Since we are concerned with neutrino masses and leptogenesis,
consider the Yukawa interactions of the left-
and right-handed leptons:
\ba
\psi_L \equiv \left(\br{c} \nu \\ e \er\right)_L \equiv (1,2,1,-1) \subset {\bf 16},\nonumber\ea
\ba 
\psi_R \equiv \left(\br{c} \nu \\ e \er\right)_R \equiv (1,1,2,-1) \subset {\bf 16}. \label{eq13}\ea
The relevant Yukawa couplings are given by
\ba
{\cal L}_Y = f \ov \psi_L \psi_R \phi + \tilde{f} \ov {\psi^c_L} \psi_L \bar
{\Delta}_L
+ \tilde{f} \ov {\psi^c_R} \psi_R \bar {\Delta}_R .\label{eq14} \ea
Then the neutrino mass matrix can be written as
\ba
M_\nu =\left(\br{cc} \nu & \nu^c \er\right)_L\left(\br{cc} m_L & m_D \\ m_D & m_R \er\right)\left(\br{c} \nu \\ \nu^c \er\right)_L,
\label{eq15}\ea

where, $m_L = \tilde{f} \langle {\bar{\Delta}_L} \rangle;~~m_R =
\tilde{f} \langle {\bar {\Delta}_R} \rangle$ and $m_D = f \langle
\phi \rangle$. Generation indices have been suppressed. The
right-handed neutrinos then remain massive, while the left-handed
neutrino masses are see-saw suppressed
\ba
m_N &=& m_R, \nonumber \\
m_\nu &=& m_L - {m_D^2 \over m_R}. \label{eq16}
\ea
The first term $m_L = \tilde{f} v_L$ is also naturally small, since
$$v_L = \langle \bar {\Delta}_L \rangle = \kappa v^2 / v_R. $$

With supersymmetry in $SO(10)$, $\kappa$ is model dependent
and some fine-tuning of this parameter is needed in
the triplet model to achieve type II see-saw dominance, successful
prediction  of large neutrino mixings and parameterization of all
fermion masses and mixings including CP-violation \cite{min1,
min2, min3, goh}. With asymptotic parity invariance in the high
scale theory, the gravitino constraint is often ignored in the
triplet model \cite{ji}.  Moreover, the observed smallness of
neutrino masses may  work against bringing the left-right symmetry
breaking scale closer to $10^{9}$ -- $10^{10}$ GeV in the
triplet model.

\par 
In Model I, we  explore an alternative approach where, without
fine-tuning of the Yukawa couplings of the see-saw formula, the
left-right symmetry breaking scale can be sufficiently lowered to
meet the requirements of resonant leptogenesis while satisfying
the gravitino constraint and maintaining consistency with
experimentally observed small values of neutrino masses.

\par
As discussed in subsequent sections, both the $SO(10)$ representations
$\bf {210}$ and $\bf {54}$ are necessary to break $SO(10) \to
{\cal {G}}_{LR}$ in Model I as well as in Model II, to  prevent
certain undesirable  scalar components of $\bf {210}$ being
lighter than the GUT scale and upsetting successful gauge coupling
unification.  

In Model I, neutrino masses arise from the Yukawa Lagrangian:
\begin{equation}
{\cal L}_Y = f \ov \psi_L \psi_R \phi + y \left( \ov\psi_L S \chi_L
+ \ov\psi_R S \chi_R \right) +  M  S^T S + H.c. \label{eqdl}
\end{equation}
where ${\chi_L}(1,2,1,-1)$ and ${\chi_R}(1,1,2,-1)$ are in the
{\bf 16} dimensional Higgs representation, $\phi$ is in a {\bf
10}, and $S$ stands for $SO(10)$ singlets, of which there are three.

The left-handed neutrinos $\nu_L$ and the right-handed 
neutrinos $N=\nu_R$ now mix with the new singlet fermions
$S$ through the mass matrix:
\ba
M_\nu =\left(\br{ccc} \nu & N^c & S \er\right)_L\left(\br{ccc} 0 & m_D & yv_L \\ m_D & 0 & yv_R \\ yv_R & yv_L & M \er\right)\left(\br{c} \nu \\ N^c \\ S \er\right)_L.
\label{eqdn}\ea
Here the Dirac neutrino mass, $m_D$, the Yukawa coupling,
$y$, and the singlet fermion mass, $M$, are $3\times 3$ matrices.
Light left-handed neutrino masses matching the experimental data
arise from this mass matrix through the double see-saw and type
III see-saw mechanisms, as has been widely discussed in the
literature \cite{min4, valle, albr}.  The model gives desired
values of neutrino masses even for low left-right symmetry
breaking scales without fine-tuning of the Yukawa couplings.

 We advance the following 
possibilities which may lead to left-right symmetry breaking at energies
much lower than in the the minimal models.
These are:
\begin{itemize}
\item[] {\sl Threshold Correction:} In the conventional analysis,
one assumes that different states within a GUT multiplet have the
same mass.  This is not exact and small splittings usually do
arise. The threshold effect due to a superheavy mass state
contributes to a small log at one-loop level; but in $SO(10)$
where big-sized representations like ${\bf 210}$ or ${\bf 126 +
\overline {126}}$ or both are used, the  one-loop contributions
by a large number of superheavy components lead to substantial
modification of the gauge couplings near the GUT scale. Both the
doublet and the triplet $SO(10)$ models belong to this
category. Thus threshold effects in each of them might
significantly change the allowed values of $M_R$ obtained from
the unification constraint.
\item[] {\sl Non-renormalizable interactions at the Planck
scale:} Since the unification scale is close to the scale of
quantum gravity, there may arise gauge invariant but
non-renormalizable interaction terms in the Lagrangian suppressed
by inverse powers of the Planck scale or a string
compactification scale. They affect the  gauge coupling values at
the GUT scale and change the predictions of the minimal models.
\item[] {\sl Additional light fields:} If there are any
additional light multiplets in the theory,
they can modify the evolution of the gauge
couplings and can allow a lowered $M_R$.
\end{itemize}
In the following, we give details of these possibilities
and show that with each of them it is possible to get
lower scale left-right symmetry breaking which in some cases 
could even be low enough to be within
striking range of the LHC/ILC.


\subsection{Threshold effects }

  Conventionally, superheavy GUT multiplets are considered to be
degenerate. In general, however, the members of a representation
could possess somewhat different masses spread around the GUT
scale giving rise to sizable  modifications of the
gauge coupling constant predictions and the mass scales
{\em via} threshold effects \cite{weinberg,hall,ovrut}.  In the
absence of precise information of the actual values of these
masses, one may assume that all the components of a particular
submultiplet are degenerate, but different submultiplets have
masses that are spread closely around the scale of symmetry
breaking \cite{hall}.  In an alternate method, one introduces a
set of effective mass parameters to capture the threshold effects
\cite{carena}.  Such an approach has been used at the SUSY $SU(5)$
scale to examine uncertainties in the GUT model predictions
\cite{langacker}. This procedure is extended here to the $\lr$
symmetry breaking scale in the form of  eq. (\ref{eqA.4})
\cite{parida1}.

Below, we examine to what extent threshold corrections
could lower the scale of left-right symmetry breaking. We assume
all superheavy gauge bosons to possess degenerate masses
identical to the unification scale $\U$.  \\

{\bf Model I:} For the particle content of Model I, from
eq. (\ref{eqA.7}) one obtains\\
\be A = B = 14/3, ~A' = 18, ~B' = 2,   ~AB' - A'B = -224/3 ,\label{eq17} \\
\ee
Using these, one has from eqs.  (\ref{eqA.5}),   (\ref{eqA.6}), and
 (\ref{eqA.8}) the following expressions for
threshold corrections on $M_R$ and $M_U$:\\
\ba
\D \ln{\R\over\Z} &=& {\pi\over 14}\left[ {10\over 3} \D'_{BL} -
8\D'_{2L} + {14 \over 3}\D'_{3C} + {25\over 3}
\D_Y - 13 \D_{2L} + {14 \over 3} \D_{3C} \right ], \nonumber \\
\D \ln{\U\over\Z} &=& {\pi\over 28}\left[ {4\over 3} \D'_{BL} +
8\D'_{2L} - {28 \over 3}\D'_{3C} +{10\over 3}\D_Y + 6\D_{2L} -
{28 \over 3} \D_{3C} \right ].  \label{eq18}
\ea
The quantities appearing on the RHS of eq. (\ref{eq18}) are
readily calculated using eq. (\ref{eqA.4}), given the superheavy
components
of $\bf {210}\op  {16} \op {\ov  {16}}\op
 {10}$. In this manner one gets  \cite{parida1},
\be  b''_{2L} = b''_{2R} = 53, ~~b''_{3C} = 56, ~~b''_{BL} = 50,
\label{eq19}\\
\ee
leading to
\ba
\D \ln{\R\over\Z} &=& {1\over 7}\left[ {125\over 3}
\ln{M_{1}\over \U} - 106 \ln{M_{2}\over \U} + {196 \over 3} \ln
{M_{3}\over \U} \right], \nonumber \\
\D \ln{\U\over\Z} &=& {1\over 7}\left[ {25 \over 3} \ln{M_{1}\over \U} +
53 \ln{M_{2}\over \U} - {196 \over 3} \ln {M_{3}\over \U}
\right]. \label{eq20}
\ea
\noindent

The pair of equations in  (\ref{eq20}) provide enough room 
to find solutions which will lead to a significant lowering of the scale $\R$
while keeping $\U$ within the Planck scale \cite{f5}.

As an illustration, one can consider a one parameter solution satisfying:
\begin{equation}
{\U\over M_1} = {\U\over M_3} = {M_2\over \U} =\eta~. 
\end{equation}

One finds from  eq. (\ref{eq20}) 
\begin{equation}
\D \ln{\R\over\Z} = - 30.42 \ln \eta, \;\;\;
\D \ln{\U\over\Z} =  15.71\ln\eta~. \label{eq21}
\end{equation}

Note that, in the absence of threshold corrections, at the
two-loop level $\ln {\U^0 \over \Z} = 33.178$ and $\ln {\R^0 \over
\Z} = 32.916$. To ensure that $\U \leq M_{Pl} =
1.2 \times 10^{19}$ GeV one must satisfy  
$\left(\D\ln{\U\over \Z}\right) \le 6.24 $. Thus, from 
eq. (\ref{eq21}) $ \eta
\le 1.48$ leading to $\left(\D\ln{\R\over \Z}\right) \ge
- 12.07$ implying
\be  \R \ge  1.0 \times 10^{11} ~~{\rm {GeV}}, ~~ \U \le 1.2 \times 10^{19}
~~{\rm {GeV}}. \label{eq22} \ee
This simple example implies that with one parameter $\eta$, $\R$
lower than that given in eq. (\ref{eq22}) corresponds to
unification scales higher than the Planck mass.  Even this bound
on $\R$ can be further lowered   by one order when smaller
threshold effects from lower scales \cite{baer, felipe} are
included leading to  $M_R \simeq 10^{10}$ GeV with near Planck
scale grand unification in the minimal doublet model. In
principle, there are three distinct mass scales $M_i, i =1,2,3,$
that enter in the threshold corrections, see  eq. (\ref{eq20}),
and there is much more flexibility to further lower $\R$. We
return to such solutions later.

It is interesting to examine how gauge coupling constants are matched
by threshold corrections to reach their common unification value 
in spite of such substantial changes in  both the mass scales.
Using  eq. (\ref{eqA.4}) and eq. (\ref{eq19}), for $ \eta
= 1.48$
the GUT-threshold corrections for individual couplings are \cite{parida1}\\
\be
\D'_{BL} = -{25\over \pi} \ln\eta = -3.16,\;\;
\D'_{2L} = {53\over 2\pi} \ln\eta = 3.35, \;\;
\D'_{3C} = -{28 \over \pi} \ln\eta = -3.54.\label{eq23}
\ee
The gauge couplings extrapolated from $ \Z$
to $\R =  10^{11}$ GeV are,\\
\be
\al_{BL}^{-1}(\R) = 53.4,  ~~\al_{2L}^{-1}(\R) = 26.3,
~~~\al_{3C}^{-1}(\R) = 18.4.\label{eq24}
\ee

With GUT-threshold effects, the one loop-evolution of the coupling constants
 from $\R$ to the new value of  $\U$,
\ba
{1\over\al_i(\U)}&=&{1\over\al_i(\R)}-{a'_i\over 2\pi}\ln{\U\over\R}+\D'_i,
 ~~~~~~ i= 2L, BL, 3C.\label{eq25}
\ea
Then using eq. (\ref{eq20}) - eq. (\ref{eq24})  in eq. (\ref{eq25}),
\be
{1\over\al_{BL}(\U)}= 23.1, \;\;\;
{1\over\al_{2L}(\U)}= 23.5, \;\;\;
{1\over\al_{3C}(\U)}= 23.7.\label{eq26}
\ee

The one parameter solution has the virtue of simplicity. However,
as noted earlier, in eq.  (\ref{eq20}) -- see also eq. (\ref{eqA.4})
--  three distinct mass scales $M_i, i =1,2,3,$ are, in general,  required to
capture the effect of the threshold corrections at the
unification scale. Table 1 depicts a whole set of such solutions.
For every solution, the effective mass splittings are within a
tolerable range and the unification scale has been increased by
the threshold corrections. The value of the unified gauge coupling
is also shown.

\begin{table}
\begin{ruledtabular}
\begin{tabular}{cccccc}
$M_R$&$M_U$& ${M_1 \over M_U}$ &${M_2 \over M_U}$ &${M_3 \over
M_U}$ &${\al_{G}^{-1}}$  \\ (GeV)&(GeV) & &&& \\ 
\hline
$ 10^{11}$&$1.2 \times 10^{19}$
&${(1.48)}^{-1}$&1.48&${(1.48)}^{-1}$&23.7 \\ 
$10^{9}$&$10^{18}$&0.272&1.770&0.831 & 23.7 \\ 
$10^{7}$&$10^{18}$&0.158&1.950&0.832 & 23.7 \\ 
$10^{7}$&$5\times10^{16}$&0.151&2.750&1.524 & 27.7 \\ 
$10^{5}$&$5\times10^{18}$&0.180&3.30&1.076 & 26.7 \\ 
$10^{3}$&$10^{19}$&0.154&4.760&1.301 & 28.7 \\ 
\end{tabular}
\end{ruledtabular}
\caption{Examples of low  intermediate scale, $M_R$, solutions
triggered by GUT-scale threshold effects in Model I (the doublet
model).}
\end{table}



\vskip0.5cm
{\bf Model II:} The threshold effect analysis for Model II  (the
triplet model) can be carried out along the same lines as in Model
I.  Thus, from eq. (\ref{eqA.7}) one finds:
\ba
 A = 38/3,~~~B = -10/3, \nonumber\\
~A' = 34,~~~~B' = -14,~~~~AB' - A'B &=& -64
.\label{eq17a} \ea
In place of eq. (\ref{eq18}) one now has
\ba
\D \ln{\R\over\Z} &=& {\pi\over 2}\left[ {8\over 9} \D'_{BL} -
3\D'_{2L} + {19 \over 9}\D'_{3C} \right ], \nonumber \\
\D \ln{\U\over\Z} &=& {\pi\over 2}\left[ {4\over 9} \D'_{BL} -
\D'_{2L} + {5 \over 9}\D'_{3C} \right ].  \label{eq18a}
\ea
The one-loop  beta-function coefficients from Model II required
for an evaluation of the RHS are:
\be  b''_{2L} = b''_{2R} = 116, ~~b''_{3C} = 122, ~~b''_{BL} = 101 .
\label{eq19a}\\
\ee
Thus, from the superheavy components of $\bf 
{210}\op  {126} \op {\ov {126}}\op  {10}$ one gets \cite{parida1}:
\ba
\D \ln{\R\over\Z} &=& \left[ {202 \over 9} \ln{M_{1}\over \U} -
{87} \ln{M_{2}\over \U} + {1159 \over 18} \ln {M_{3}\over
\U} \right], \nonumber \\
\D \ln{\U\over\Z} &=& \left[ {101 \over 9} \ln{M_{1}\over \U}
-{29} \ln{M_{2}\over \U} + {305 \over 18} \ln {M_{3}\over
\U} \right].~~~~~~\label{eq20a}
\ea
\noindent

Eqs. (\ref{eq20a}) depend, as in the case of Model I, on the three
mass scales $M_i, i=1,2,3$ which can be chosen appropriately to
ensure a solution with a low intermediate scale $\R$. A few
typical examples are presented in Table 2. It is noteworthy that
the gauge coupling at unification is larger for these solutions
than for the ones in Table 1. 

\begin{table}
\begin{ruledtabular}
\begin{tabular}{cccccc}
$M_R$&$M_U$& ${M_1 \over M_U}$ &${M_2 \over M_U}$ &${M_3 \over
M_U}$ &${\al_{G}^{-1}}$  \\ (GeV)&(GeV) & &&& \\ \hline
$5\times 10^{9}$&$1.58\times10^{16}$ &2.204 &1.200 &0.659 &15.0 \\ 
$10^{10}$&$1.58\times10^{16}$ &2.065&1.160 &0.659 & 15.0 \\ 
$10^{11}$&$1.58\times10^{16}$ &1.661&1.050 &0.656 & 15.0 \\
\end{tabular}
\end{ruledtabular}
\caption{Examples of low  intermediate scale, $M_R$, solutions
triggered by GUT-scale threshold effects in Model II (the triplet
model).}
\end{table}


  Before moving on, let us remark that in many of the threshold
effect driven solutions in Model I the unification scale is
pushed to higher values. It is well known that suppression of
Higgsino mediated supersymmetric proton decay modes like $p \to
K^+ {\ov {\nu}}$, $p \to K^0 {\mu}^+$ etc. is a generic problem
in minimal SUSY GUTs and the amplitudes are proportional to
$\U^{-2}$.  The higher unification scales help to evade this
problem in a natural and effective fashion with a suppression
factor $({\U^0\over \U})^2 = 10^{-2} - 10^{-4}$.

\subsection{\bf  Planck scale effects}

Since the GUT scale is close to the Planck mass, it is possible
that gravity induced non-renormalizable terms could change the
usual field theoretic predictions of gauge coupling
unification. These interactions are suppressed by inverse powers
of the Planck mass.  For example, consider the gauge invariant
 Lagrangian consisting of the $\rm {dim}.5$
non-renormalizable operators (NRO),
\ba
{\cal L}_{NRO}&=& -{\eta_1\over
2M_G}Tr\left(F_{\mu\nu}\Phi_{210}F^{\mu\nu}\right) 
-{\eta_2\over
2M_G} Tr\left(F_{\mu\nu}\Phi_{54}F^{\mu\nu}\right).
\label{eq27}
\ea
The effective gauge coupling constants at the unification point
get changed due to these non-renormalizable terms.
In particular, these interactions determine the
parameters in eq.   (\ref{eq13a}) and one finds  \cite{shafi,
parpat},
\ba\ep_{2L}&=&\ep_{2R}=-{3\over 2}\ep_2,
~\ep_{3C}= \ep_2-\ep_1,~~\ep_{BL}=2\ep_1+\ep_2,\nonumber\\
\ep'&=& {4\over 3}\ep_1 + {5 \over 3}\ep_2, ~~ \ep'' = 4\ep_1 - 5\ep_2,
\nonumber
\ea
where
\be 
\ep_1={{3\eta_1}\over 4}{\U\over M_{G}}\left[{1\over
{4\pi\al_G}}\right]^{1\over 2},
\;\;\;
 \ep_2={{3\eta_2}\over 4}{\U\over M_{G}}\left[{1\over
{15\pi\al_G}}\right]^{1\over 2},\label{eq28}
\ee
leading to the following analytic expressions for the corrections
on the mass scales,
\ba\left(\D\ln{\R\over \Z}\right)_{gr}&=&{2\pi(A'\ep'-A\ep'')\over\al_G
(AB'-A'B)}, 
= -{\pi\over {7\al_G}}\left[\ep_1 + 10\ep_2\right],\nonumber \\
\left(\D\ln{\U\over \Z}\right)_{gr}&=&{2\pi(B\ep''-B'\ep')\over\al_G(AB'-A'B)}
= {\pi\over{7\al_G}}\left[5\ep_2-3\ep_1\right].~~~~~~~~\label{eq29}
\ea
While the change in the mass scales are governed by the
above relations the individual coupling constants near the
GUT scale change as,\\
\be
 \D_{2L}^{\rm {(gr)}} = {3\ep_2\over 2\al_G}, \;\;\;
 \D_{BL}^{\rm {(gr)}} = -{\left(2\ep_1+\ep_2\right)\over \al_G}, \;\;\;
 \D_{3C}^{\rm {(gr)}} =  {\left(\ep_1-\ep_2\right)\over \al_G}. \label{eq30}
\ee
Using the most natural scale for the  two NRO's as the Planck
mass, $M_G = 1.2\times 10^{19}$  GeV,  and eq. (\ref{eq28}) -
eq. (\ref{eq30}) we searched for gravity corrected solutions for
low intermediate mass scale and high GUT scale with the
constraint $|\eta_{1,2}| \simeq O(1)$.

For example with $\epsilon_1 = 0.15$, $\epsilon_2 = 0.174$, $M_G
= M_{Pl.}$  we have   $M_R = 10^7$ GeV and $M_U = 10^{18.4}$ GeV,
corresponding to $\eta_1 = 0.494$ and $\eta_2 = 1.160$.  The
corrections to the coupling constants are obtained through
$\D_{BL}^{\rm {(gr)}} = - 11.47$, $\D_{2L}^{\rm {(gr)}} =  6.52$,
and $\D_{3C}^{\rm {(gr)}} = 0.6$ .  When these are added to
one-loop extrapolated values from $M_Z$ to $M_U ~(\equiv
10^{18.4}$ GeV), the three coupling constants match consistently
with their common value  $\al_G^{-1} \simeq 25$.  All solutions
with high unification scales require $|\eta_{1,2}| \simeq O(1)$
as shown in Table 3. Thus, $\rm {dim.}5$ operators are capable of
lowering the left-right symmetry breaking scale to  $\R = 10^5 -
10^9$ GeV, making Model I consistent with large neutrino mixing
and leptogenesis when the minimal doublet model is supplemented
by the addition of a ${\bf 54}$.

We find that  
high values of $\U \simeq 10^{18}$ GeV  require smaller 
$\eta_{1,2} \simeq O(1)$ while a lower  $\U \simeq 10^{16}$ GeV  
requires unnaturally  larger values of the parameters. The preferred 
solutions with naturally large values of $\U$ exhibit the virtue of 
suppression of Higgsino mediated proton decay by factors $({\U^0\over\U})^2 =
10^{-3} - 10^{-4}$.

We now extend the triplet model by the addition of a Higgs representation 
{\bf 54} and including the effects of the two non-renormalizable operators
of eq. (\ref{eq27}). The changes in the mass scales are given by
\ba\left(\D\ln{\R\over \Z}\right)_{gr}&=& 
 -{\pi\over {12\al_G}}\left[-2\ep_1 + 45\ep_2\right],\nonumber \\
\left(\D\ln{\U\over \Z}\right)_{gr}&=&
 -{\pi\over{12\al_G}}\left[2\ep_1+15\ep_2\right].\label{eq31}
\ea

Unlike for  the doublet model, we find that gravitational corrections alone 
do not succeed in substantially reducing the $M_R$ scale. This
behaviour of the triplet model can be understood in terms of the
larger Higgs representations -- {\bf 126} and ${\bf \overline{126}}$
--  involved and the consequent tension with perturbativity (see
Sec.\ref{sec:lb}).

\begin{table}
\begin{ruledtabular}
\begin{tabular}{ccccc}
$M_R$&$M_U$&  $\eta_1$ &$\eta_2$&${\al_{G}^{-1}}$ \\
(GeV)&(GeV) & && \\ \hline
$10^{9}$&$3.16\times 10^{18}$&0.305&0.96&25.00 \\
$10^{7}$&$3.16\times 10^{18}$&0.494&1.16&25.64 \\
$10^{6}$&$8\times 10^{17}$&2.728&4.77&25.32 \\ 
$10^{5}$&$3.16\times 10^{18}$&0.671&1.34&25.32 \\ 
\end{tabular}
\end{ruledtabular}
\caption{Sample solutions with low intermediate scales, $M_R$,
obtained for Model I (the doublet model) through Planck scale
induced interactions parameterized by $\eta_1$ and $\eta_2$ (see
text).}
\end{table}


\subsection{Doublet model with additional light multiplets}

The third and final alternative that we discuss for obtaining a
low intermediate scale in Model I is through additional light
chiral submultiplets. We find that if there are appropriate light
states in the particle spectrum then the unification of gauge
couplings is consistent with a significant lowering of $\R$.

In earlier work  attempts have been made to obtain
intermediate scales much lower than the GUT scale by
spontaneous breaking of SUSY $SO(10)$ in the first step and the
gauge group $\lr$ in the second step with or without \cite{lee}
left-right discrete symmetry.  The crucial point of this
paper is that we require the  left-right symmetric gauge group
with $ g_{2L} = g_{2R}$ to survive to low intermediate scales in
order to evade the gravitino problem and at the same time obtain
low mass $W_R^{\pm}$ gauge bosons to possibly even provide
testable signals at collider energies in the near future.

We present below two models which meet these requirements. The
models are identical up till the scale $M_R$ and consist of the
MSSM particles.  They differ in the number and type of
additional chiral multiplets which contribute in the range  $
M_R$ to $M_U $. In this subsection, we choose to distinguish
between the SUSY scale, $M_S$ (which is chosen at 1 TeV),  and
$M_Z$. The RG evolution of the couplings from $M_Z$ to $M_S$ is
governed by the  one- and two-loop coefficients:
\ba
&&\left(\br{c} a_{Y}\\ a_{2L}\\ a_{3C}\er\right)=\left(\br{c} {21\over 5}\\
-3\\ -7\er\right), ~~
b_{ij}=\left(\br{ccc} {104\over 25}&{18\over 5}&{44\over 5}\\ {6\over 5}&8&
12\\
{11\over 10}&{9 \over 2}&-26\er\right),\, i,j \subset \std, \label{eq10b}
\ea
while from $M_S$ to the scale $M_R$ eq. (\ref{eq10}) 
is applicable. In eq.(\ref{eq10b}) the beta-function coefficients have been 
derived assuming two light doublets in the nonSUSY model below $M_S$ which 
 emerges naturally from the MSSM existing above $M_S$. \\

\par\noindent
{\bf Model A:}
  In addition to the  MSSM particles, we assume that
supermultiplets with  
the following gauge quantum numbers are light
with masses at the $M_R$ scale:
\ba
\sigma (3, 1, 1, 4/3)\op {\ov {\sigma}}(\ov 3, 1, 1, -4/3)
& \subset & {\bf 45, 210}, \nonumber\\
\eta(1, 1, 1, 2) \op {\ov {\eta}}(1, 1, 1, -2) & \subset & {\bf 120}.\label{eq33}
\ea
\par\noindent
The one- and two-loop
coefficients including these fields are,
\ba
&&\left(\br{c} a'_{BL}\\ a'_{2L}\\ a'_{2R}\\ a'_{3C}\er\right)
=\left(\br{c} 16\\ 2\\ 2\\ -2\er\right),\label{eq34}
\ea
\ba
&& b'_{ij}=\left(\br{cccc} {241/6}&{27/2}&{27/2}&{88/3}\\{9/2}&{32}&{3}&
{24}\\{9/2}&{3}&{32}&{24}\\{11/3}&{9}&{9}&{76/3}\er\right),\,
 i, j=BL, 2L, 2R, 3C.\label{eq35}\ea

\begin{figure}[htbp]
  \centering
  \includegraphics[width=0.85\textwidth,height=0.25\textheight]
  {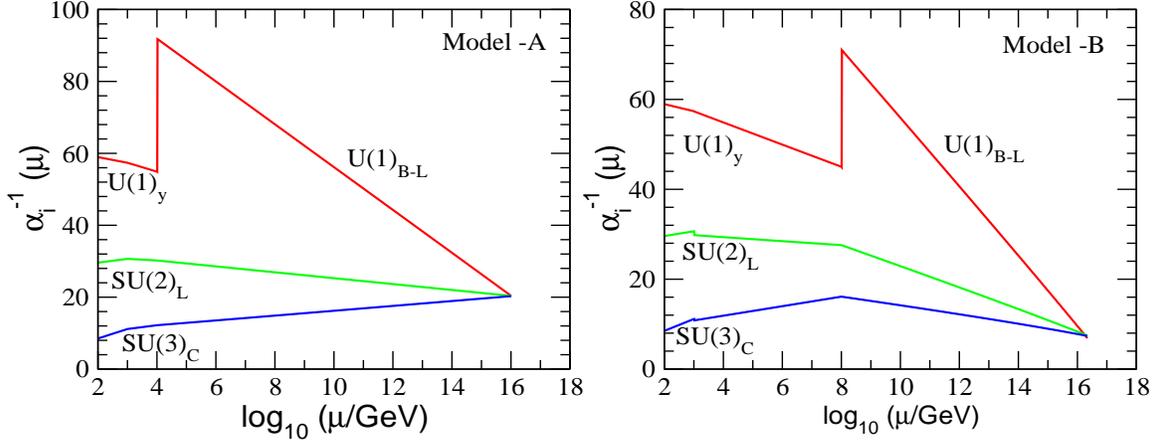}
   \caption{The evolution of the gauge couplings in
models with additional light multiplets. The left- (right-) panel
corresponds to Model A (Model B).}
  \label{f:coupl}
\end{figure}

At two-loop level the evolution of gauge couplings and their
unification have been shown in Fig. \ref{f:coupl} for $M_R = 10^4$ GeV.
Some sample solutions to the RGEs for gauge couplings with allowed values of
$\R$, $\U$ and the GUT fine structure  constant ($\al_G$) are
presented in Table 4.  We find that with the grand unification
scale $M_U = 2\times 10^{16}$ GeV, an intermediate scale in the
range of $M_R = 5$ TeV - $10^{10}$ GeV is possible in this model
with excellent unification of the gauge couplings. In spite of
the presence of additional fields, the gauge couplings at the GUT
scale remain perturbative in a manner similar to the minimal GUT
with $\alpha_G^{-1} = 22.22 - 20.40 $.
\vskip0.1cm
\par\noindent
{\bf Model B:}
In addition to the MSSM particles we assume
that there are additional superfields  with their masses at the
$M_R$ scale which transform as:
\ba
\xi (6,1, 1, 4/3)\op {\ov {\xi}}(\ov 6,1, 1, -4/3, )  & \subset &
{\bf 54}, \nonumber\\
\eta(1, 1, 1, 2) \op {\ov {\eta}}(1,1, 1,-2)  & \subset & {\bf 120},\nonumber\\
C(1,2,2,0)  & \subset & {\bf 10, 120, 126}, \nonumber \\ 
 D_L(1,3,1,0)\op D_R(1,1,3,0)  & \subset & {\bf 45, 210},\label{eq36}
\ea
where we have used a pair of $C(1,2, 2, 0)$.
\par\noindent
The one- and two-loop coefficients in this scenario are
\ba
&&\left(\br{c} a'_{BL}\\ a'_{2L}\\ a'_{2R}\\ a'_{3C}\er\right)
=\left(\br{c} 20\\ 6\\ 6\\ 2\er\right),\label{eq37}
\ea
\ba
b'_{ij}=\left(\br{cccc} {305/6}&{27/2}&{27/2}&{344/3}\\{9/2}&{70}&{9}&
{24}\\{9/2}&{9}&{70}&{24}\\{43/3}&{9}&{9}&{332/3}\er\right), i, j=BL, 2L, 2R, 3C. \label{eq38}\ea

Gauge coupling evolution and unification in this case is shown in
Fig. \ref{f:coupl} for an example with $\R = 10^{8}$ GeV. A
couple of sample solutions with $\R$ which satisfy the gravitino
constraint are presented in Table 4. For this alternative, the
intermediate scales are typically in the range of $M_R = 10^7$
GeV - $10^{10}$ GeV. A very precise unification of the gauge
couplings has been found  when further small SUSY threshold
effects at the TeV scale are taken into account
\cite{baer}. Because of these  effects, the resulting gauge couplings show 
small discontinuities
at $M_S = 10^3$ GeV as shown in the Fig. \ref{f:coupl} for Model B. The gauge
couplings near the GUT scale approach strong coupling 
($ \alpha_G \simeq 0.1 $) as shown in Table 4 and Fig.
\ref{f:coupl}.

\begin{table}
\begin{ruledtabular}
\begin{tabular}{lccr}
Model&${\rm M_R}$&${\rm M_U}$&${\al_{G}^{-1}}$  \\
&(GeV)&(GeV)& \\ \hline
&$10^{9}$&$1.15\times 10^{16}$&22.22\\
A&$10^{5}$&$1.10\times 10^{16}$&20.83\\
&$10^{4}$&$ 10^{16}$&20.40\\\hline
&$10^{9}$&$1.82\times 10^{16}$&7.58\\
B&$10^{8}$&$2.00\times 10^{16}$&10.13\\ 
\end{tabular}
\end{ruledtabular}
\caption{Sample solutions for low intermediate scales, $\R$, in two models with additional light multiplets at the intermediate
scale (see text).}
\end{table}

We show in the next section that the intermediate scale in the
triplet model has a lower bound at $10^9$ GeV which is expected
to be increased by  additional Higgs scalars at the $M_R$ scale.

From the above two examples and earlier investigations it is
clear that right-handed mass scales as low as $M_R= 5$ TeV $-
10^{10}$ GeV are viable when additional light chiral multiplets at
the $M_R$ scale are admitted. As already noted, such low scales
are necessary for the successful implementation of leptogenesis
in the doublet model (Model I). Obviously, these models may have
interesting new signatures at LHC and future collider
experiments.  It is noteworthy that all the light multiplets
exploited in the two models are contained in $SO(10)$
representations which have been invoked in the literature anyway
for various purposes.


\section{\bf Lower bound on intermediate scale in the triplet model}
\label{sec:lb}

As pointed out earlier, the  high dimensional Higgs
representations like ${\bf {210}}$ and/or ${\bf {126} + \ov
{126}}$ result in large threshold corrections at the GUT scale
even if their superheavy components are only few times heavier or
lighter than $\U$. In this respect, threshold corrections in the
triplet model with ${\bf {126} + \ov {126}}$ are more significant
compared to those in the doublet model which uses ${\bf {16}+ \ov
{16}}$. Normally, one would therefore expect to obtain lower
$M_R$ in the former model.

In this section we show that this is not true and, in fact,
establish that $M_R$ cannot be lower than $10^9$ GeV in the
triplet model. This lower bound is set by the perturbative
renormalization group constraint when parity survives in the
left-right gauge group as happens in the case of $\cal {G}_{LR}$.
As the GUT threshold effects contribute only at the unification
scale, we use the two-loop equation for $\al_{BL}$ between $M_R$
and $ M_U$ with the corresponding coefficients given in eq.
(\ref{eq10}) and eq. (\ref{eq11a}).  It is seen that if $M_R \le
10^{9}$ GeV, $\al_{BL}$ exceeds the perturbative limit ($\simeq
1$) before the GUT scale is reached.

Analytically, this  behavior of the gauge coupling becomes transparent 
by noting that 
the position of the Landau pole ($\mu_0$), where $g_{BL}(\mu_0) = \infty$,
is given by,
  
\ba
\mu_0 &=& M_R ~\exp\left[{2\pi\over a'_{BL}}{1\over \al_{BL}(M_R)}\right] .
\label{eq38a}
\ea

Here
\ba
{1\over \al_{BL}(M_R)}= {5\over 2}\left({1\over
\al_Y(M_Z)}-\Theta_Y + \D_Y\right)\nonumber ~~~~~~~~~~~~~~~~~~~~~~~\\ 
-{3\over2}\left({1\over\al_{2L}(M_Z)}-\Theta_{2L}+\D_{2L}\right)
- {1\over 4\pi}\left(5a_Y - 3a_{2L}\right)\ln {M_R
\over M_Z}.\nonumber \\ 
\label{eq39} \ea

Using  eq. (\ref{eq39}) we calculate $\al_{BL}^{-1}(M_R)$ for
$M_R = 10^3$ GeV to $10^{11}$ GeV from low energy data  ignoring
the small threshold effect due to superpartners and use them in
eq. (\ref{eq38a}) to estimate the value of $\mu_0$. Our two-loop
estimations of the pole position are shown in Table 5 for the
triplet model with $a'_{BL}=24$.  The two-loop corrections
predict slightly lower values of $\mu_0$ than eq. (\ref{eq38a}).
For intermediate scales $M_R = 1$ TeV  to  $10^9$  GeV, the pole
positions are found  in the range $7.76\times 10^{13}$ GeV to
$1.44\times 10^{16}$ GeV indicating that for the $U(1)_{BL}$
gauge coupling perturbation theory breaks down below the GUT
scale for these values of $M_R$.  When $M_R \gsim 10^{10}$ GeV,
the pole positions occur clearly above the GUT scale with $\mu_0
\gsim  3.46\times 10^{16}$ GeV.   In other words, with only the
minimal particle content  needed to maintain supersymmetry and
left-right symmetry below the GUT scale, from the requirement of
perturbativity the triplet model leads
to the conservative lower bond on the intermediate scale,
\be
M_R > 10^9 ~ {\rm GeV}. 
\ee

 Inclusion of additional new scalar degrees of freedom anywhere
between $M_R$ to $M_U$ would increase the one-loop
beta-function coefficient of the $U(1)_{B-L}$ gauge coupling and
bring down the pole position further. This, in turn, would
further tighten the lower bound on $M_R$ beyond $10^{9}$ GeV.
This is why, unlike in the doublet model, the presence of additional
light scalars near $M_R$ cannot reduce the value of the
intermediate scale in the triplet model.

In contrast to the triplet model for which $a'_{BL}=24$, the
doublet model has $a'_{BL}=9$  which enhances the argument of
the exponential on the RHS of eq. (\ref{eq38a}) by a factor
$\simeq 24/9 = 2.66$ compared to the triplet model for the same
value of $M_R$. Such a factor in the argument pushes the Landau
pole to a position much above the GUT scale. Thus, even with $M_R
= 1$ TeV, whereas the triplet model pole position is at
$\mu_0\simeq 1.18\times 10^{14}$ GeV which is approximately two
orders below the GUT scale, in the the doublet model the pole
occurs at $\mu_0 \simeq  3.3\times 10^{32}$ GeV.  Although this
latter scale for the doublet model is expected to be
substantially lower because of the contribution of superheavy
particles near the GUT scale, it is clear that the coupling
constant never hits a Landau pole below the
GUT-Planck  scales $\simeq 10^{18}$ GeV. This tallies with the
results in Sec.III  where solutions have
been obtained using threshold and gravitational corrections.

\begin{table}
\begin{ruledtabular}
\begin{tabular}{lcr}
$M_R$&${1\over \al_{BL}(M_R)}$&$\mu_0$\\
(GeV)&&(GeV) \\\hline
$10^{3}$&$97.429$&$7.76\times 10^{13}$\\
$10^{5}$&$86.407$&$4.56\times 10^{14}$\\
$10^{7}$&$75.406$&$2.56\times 10^{15}$\\
$10^{8}$&$69.907$&$6.16\times 10^{15}$\\
$10^{9}$&$64.409$&$1.44\times 10^{16}$\\
$10^{10}$&$58.912$&$3.46\times10^{16}$\\
$10^{11}$&$53.415$&$8.31\times10^{16}$\\
\end{tabular}
\end{ruledtabular}
\caption{Location of Landau poles, $\mu_0$, signifying violation of
perturbativity, for different choices of the 
intermediate scale $M_R$ in the triplet model.}
\end{table}


With such a lower bound on $M_R$ in the triplet model, this
version of  SUSY $SO(10)$ rightly deserves its description as a high
scale theory. The SUSY $SO(10)$ triplet model appears to fit
ideally for description of quark-lepton masses and mixings
through high-scale $b-\tau$ unification and type II see-saw
dominance or even through  type I see-saw mechanism~\cite{min2, goh, babu} .

Since $M_R \gsim 10^{9}$ GeV in the triplet model, the lightest
right-handed neutrino mass could satisfy the gravitino
constraint, but in this case generating the quark and lepton
masses and mixings  has to be re-examined.  While a detailed
analysis of neutrino data is yet to emerge in the doublet model,
it is well known that reproducing small neutrino masses is no
problem even if the right-handed neutrinos are near the TeV
scale.  With such a low value of $M_R$  the desired criteria of TeV
scale resonant leptogenesis is fulfilled and through the
$W_R^{\pm}$ and  $Z_R$ bosons and the light Higgs scalars,
$\chi_L^{\pm}, \chi_L^{0}$, $\chi_R^{\pm}$, and $ \chi_R^{0}$,
the model can be tested  at the LHC and ILC.
The superpartner of the lightest right-handed neutrino in the
doublet model may also be a good candidate for dark matter.

\section{\bf Remarks on light scalars and fermion masses in minimal $SO(10)$}

One of the most appealing features of the minimal supersymmetric
$SO(10)$ model is that one can calculate the pattern of symmetry
breaking and predict fermion mass relations at the GUT scale
\cite{fmsen}. Concomitant with these, in the minimal model, is
an intermediate left-right breaking scale, $\R$, constrained to
be rather close to the GUT scale $\U$.  Can the virtues of the
model be made to survive when $M_R$ is lowered?

Let us briefly summarize the salient features with reference to
Model II. The Higgs fields are: $$ \Phi \equiv {\bf 210}, ~~~~~
\Sigma \equiv {\bf 126}, ~~~~~ \bar \Sigma \equiv {\bf
\overline{126}}, ~~~~~ H \equiv {\bf 10}, $$ where $\Delta_{L,R}
\subset \Sigma$ and $\phi \subset H$. The fermions belong to the
representation $\Psi \equiv$ {\bf 16}.  The complete
superpotential of the model can then be written as
\begin{equation}
W = W_Y + W_H,
\end{equation}
where the Yukawa couplings are in $W_Y$ and the scalar potential
can be derived from $W_H$. They can be written as (we follow the
notations of ref. \cite{min3})
\ba
W_Y &=& Y_{10} \Psi \Psi H + Y_{126} \Psi \Psi \bar \Sigma  ,
\nonumber \\
W_H &=& {m_{\Phi} \over 4!} \Phi \Phi + {\lambda \over 4!} \Phi \Phi \Phi
+ {M \over 5!} \Sigma \bar \Sigma + {\eta \over 4!} \Phi \Sigma
\bar \Sigma \nonumber \\
&& + m_H H H + { 1 \over 4!} \Phi H ~(\alpha \Sigma + \beta \bar \Sigma).
\label{eq47}
\ea
As usual, minimization of the scalar potential  gives the
allowed values of the $vev$ of the different fields. In addition,
fermion mass relations are also determined in terms of the
parameters of the model.

 It may appear 
that the solutions presented earlier with lowered left-right symmetry 
breaking scales are in conflict with results on fermion masses.
However, this need not be the case. 
For example, when gravitational corrections are
included, there may well be non-renormalizable terms in the
superpotential, suppressed by the Planck scale, which can
contribute to the Yukawa couplings after the GUT symmetry 
breaking by the field $\Phi$. 
Thus, in the presence of such corrections, the
superpotential will have to be supplemented by
\begin{equation}
W_Y^{G} = {1 \over M_{Pl}} (Y^G_{10} \Psi \Psi H \Phi + 
Y^G_{126} \Psi \Psi \bar{\Sigma} \Phi) + \cdots .
\end{equation}
These new interactions will be suppressed by ${\langle \Phi
\rangle / M_{Pl}}$. But  $\langle \Phi \rangle \sim \U$ is close to
the Planck scale, as we have illustrated, and hence the
suppression need not be too much. In addition, the non-renormalizable
couplings $Y^G$ could also be large. Then the fermion mass
relations obtained for the minimal supersymmetric $SO(10)$ models
could be radically affected.  Fermion mass relations can also get
changed in the presence of new Higgs scalars. Thus the
low intermediate mass scales, $M_R$, obtained in the present
analysis need not be inconsistent with the fermion mass
relations.

At the tree level, the minimal triplet
model predicts ~\cite{min3}  masses near
$M_R$ for  additional states belonging to ${\bf 210}$ with the
quantum numbers
\ba
E_L(3, 3, 1, 4/3)\op {\ov E}_L(\ov 3, 3, 1, -4/3), \nonumber\\
E_R(3, 1, 3, 4/3) \op {\ov E}_R(\ov 3, 1, 3, -4/3).\label{eq48}
\ea

We have checked that with the minimal Higgs content, the
renormalizable doublet model also leads to similar light Higgs
scalars. It has been further  noted in ref. \cite{min3} that
these states prevent having  parity conserving ${\cal G}_{LR}$ at
any value of the  intermediate scale below $M_U$.  We remark that
their presence at $M_R$ sufficiently lower than $M_U$, apart from
being in conflict  with  $\sin^2\theta_W(M_Z)$ and $\al_S(M_Z)$,
spoils perturbative gauge coupling evolutions by developing
Landau poles in the coupling constants in the region $M_R < \mu <
M_U$.  This difficulty could be  avoided \cite{f6} by extensions of the
minimal doublet or the triplet model through the inclusion of
non-renormalizable operators and/or additional $SO(10)$ Higgs
representations,  like $\bf {54}$. For example, the presence of the
non-renormalizable term in the superpotential
\ba
W_{gr} &=& {\lambda^G \over {4!M_G}} \Phi^4, \nonumber  
\ea 
with $M_G = M_{Pl}$, or  (string) compactification scale, can
lift the masses of these light scalars close to the GUT scale
when the ${\bf 210}$ gets $vev$ along the direction
$\langle\Phi^0\{15,1,1\}\rangle~\sim~M_U$, leading to $M_E
={{2\lambda^G m_{\Phi}^2}/{\lambda^2 {M_G}}}$. Then their
contributions are added to GUT-threshold effects,  as discussed
earlier.

\section{Summary and conclusion}

In this work, we have discussed the question of low intermediate
left-right symmetry breaking scales, as preferred by
leptogenesis, in the minimal supersymmetric $SO(10)$ GUTs with
only doublet Higgs scalars as well as with triplet scalars.
In view of the presence of additional scalar components predicted
from mass spectra analysis \cite{min3} which disrupt perturbativity and 
gauge coupling
unification, the minimal renormalizable triplet model with Higgs 
representations
${\bf 210 \oplus 126 \oplus {\ov {126}} \oplus 10}$ is ruled out as a 
candidate 
for any value of left-right symmetry breaking intermediate scale. 
With the added presence of a Higgs representation $54$ and/or 
non-renormalizable interactions, these unwanted scalar components are
made superheavy and 
we find, in agreement with previous work, that in the minimal
models, at the one-loop level gauge coupling unification requires
the scale of left-right symmetry breaking to be close to the GUT
scale \cite{f7}. Inclusion
of the two-loop contributions  eliminates
even this possibility as no solution can be found at all with an
intermediate scale.  On the other hand, evading the gravitino
problem, which would otherwise plague successful big bang
nucleosynthesis, would require $M_R \leq 10^9~ {\rm GeV}$.  We
have pointed  out that this impasse can be circumvented in the
case of the doublet model by including threshold corrections near
the GUT scale, including non-renormalizable interactions due to
gravity induced Planck scale effects, or by adding new light
scalar multiplets.  In the last alternative, the additional light
submultiplets used are present in representations commonly used in
$SO(10)$ non-minimal models, but they are different from those which emerge
from mass spectra analysis~\cite{min3}.  These considerations
allow the left-right symmetry breaking scale to be low, as low as
even a few TeV, making it phenomenologically interesting.  The
unification scale obtained in the doublet model using the first two methods 
turns out to be large,
making it safe for Higgsino mediated proton decay as well as with
fermion mass relations.  
In the triplet model, although threshold
effects can easily decrease the intermediate scale, we find a
perturbative lower bound, $M_R > 10^9$ GeV, below which the
intermediate scale cannot be lowered. With this bound, the
triplet model with an added ${54}$ and/or nonrenormalizable interactions
 emerges as a high scale theory of SUSY $SO(10)$
description of fermion masses and mixings. In this model the
possibility of meeting the gravitino constraint can be fulfilled
provided neutrino masses and mixings  are successfully reproduced
with $M_R \gsim 10^{9}$ GeV.  With $M_R$ in the TeV region in the
doublet model, apart from successful resonant leptogenesis with
full compliance of the gravitino constraint, the model
predictions  can be tested through their various manifestations
\vspace{0.3cm}
at the LHC and ILC.\\
\vspace{.2cm}
\hspace{7cm}{\bf ACKNOWLEDGMENTS} \\
M. K. P. thanks Harish-Chandra Research Institute, Allahabad, India for 
hospitality and Institute of Physics, Bhubaneswar, India for facilities.


\begin{thebibliography}{[99]}

\bibitem{lr} J. C. Pati and A. Salam, Phys. Rev. {\bf D10} (1974)
275; R. N. Mohapatra and J. C. Pati, Phys. Rev. {\bf D11} (1975)
566; R. N. Mohapatra and J. C. Pati, Phys. Rev. {\bf D11} (1975) 2558; G. Senjanovi\'c and R. N. Mohapatra, Phys. Rev. {\bf
D12} (1975) 1502.

\bibitem{dpar} D. Chang, R. N. Mohapatra and M. K. Parida, Phys.
Rev. Lett. {\bf 52} (1984) 1072; Phys. Rev.  {\bf D30} (1984)
1052; D. Chang, R. N. Mohapatra, J. M. Gipson, R. E. Marshak and M.
K. Parida, Phys. Rev.  {\bf D31} (1985) 1718.

\bibitem{so10} H. Georgi, in {\em Particles and Fields -- 1974},
ed. C. A. Carlson (AIP, New York, 1975); H. Fritzsch and P.
Minkowski, Ann. Phys. (N.Y.) {\bf 93} (1975) 193; T. Clark, T. Kuo
and N. Nakagawa, Phys. Lett. {\bf B115} (1982) 26; C. S. Aulakh
and R. N. Mohapatra, Phys. Rev. {\bf D28} (1983) 217.

\bibitem{min1}K. S. Babu and R. N. Mohapatra, Phys. Rev. Lett. {\bf 70}
(1993) 2845.

\bibitem{min2} B. Bajc, G. Senjanovi\'c and F. Vissani,
 Phys. Rev. Lett. {\bf 90} (2003) 051802;
H. S. Goh, R. N. Mohapatra and S. P. Ng,
Phys. Lett. {\bf B570} (2003) 215; Phys. Rev. {\bf D68}
(2003) 115008; 
S. Bertolini, M. Frigerio and M. Malinsky, 
Phys. Rev. {\bf D70} (2004) 095002.

\bibitem{min3}  
B. Bajc, A. Melfo, G. Senjanovi\' c and F. Vissani,
Phys. Rev. {\bf D70} (2004) 035007; 
C. S. Aulakh, B. Bajc, A. Melfo, G. Senjanovi\'c, and F. Vissani,
Phys. Lett. {\bf B588} (2004) 196; 
T. Fukuyama, A. Ilakovic, T. Kikuchi, S. Meljanac and
N. Okada, J. Math. Phys. {\bf 46} (2005) 033505; Eur. Phys. J. 
{\bf C42} (2005) 191.


\bibitem{min4} S. M. Barr, Phys. Rev. Lett. {\bf 92} (2004)
101601; U. Sarkar, Phys. Lett. {\bf B622} (2005) 118;  K. S.
Babu, I. Gogoladze, P. Nath, and R. M. Syed, Phys. Rev. {\bf D72}
(2005) 095011.

\bibitem{lept}  
S. Davidson and A. Ibarra,  Phys. Lett. {\bf B535} (2002) 25: T.
Hambye and G. Senjanovi\'c, Phys. Lett. {\bf B582} (2004) 73; G.
D'Ambrosio, T. Hambye, A. Hektor, M. Raidal and A. Rossi, Phys.
Lett. {\bf B604} (2004) 199; L. Boubekeur, T. Hambye and G.
Senjanovi\'c, Phys. Rev. Lett. {\bf 93} (2004) 111601; N. Sahu
and S. Uma Sankar,  Phys. Rev. {\bf D71} (2005) 013006.

\bibitem{lepto}  M. Flanz, E. A. Paschos, and U. Sarkar, Phys. Lett.
 {\bf B345} (1995) 248; Phys. Lett. {\bf B389} (1996) 69; A. Pilaftsis, 
Nucl. Phys. {\bf B692} (2004) 303; W. Buchmuller and M. Plumacher, Phys. 
Lett. {\bf B431} (1998) 354; W. Buchmuller, P. Di Bari, and M.
Plumacher, New J. Phys. {\bf 6} (2004) 105.

\bibitem{f1} Electric charge is normalized in terms of the $SU(2)_L$, $SU(2)_R$, $U(1)_{(B-L)}$ and $U(1)_Y$ quantum numbers as $ Q = T_{3L} + T_{3R} + {B - L \over 2} = T_{3L} + {Y \over 2}$.




\bibitem{gj} H. Georgi and C. Jarlskog, Phys. Lett. {\bf B86} (1979)
297.


\bibitem{valle} R. N. Mohapatra, Phys. Rev. Lett. {\bf 56} (1986) 561;
R. N. Mohapatra and J. W. F. Valle, Phys. Rev. {\bf D34} (1986) 1642.


\bibitem{new} U. Sarkar, ref.\cite{min4}.

\bibitem{see-saw} P. Minkowski, Phys. Lett. {\bf B67} (1977)
421; M. Gell-Mann, P. Ramond and R. Slansky, in {\it
Supergravity}, eds. D. Freedman {\it et al.} (North-Holland, Amsterdam,
1980); T. Yanagida, in proc. KEK workshop, 1979
(unpublished); R. N. Mohapatra and G. Senjanovi\'c,
Phys. Rev. Lett. {\bf 44} (1980) 912; S. L. Glashow, {\it Carg\`ese
lectures}, (1979).


\bibitem{goh} H. S. Goh, R. N. Mohapatra and S. Nasri, Phys. Rev. {\bf D70}
  (2004) 075022; K. S. Babu and C. Macesanu, Phys. Rev. {\bf D72} (2005) 
  115003.

\bibitem{f2}Here, for simplicity of discussion, we have considered 
$\tilde{f}$ to be multiplying a unit matrix in flavor space. The RH
neutrino masses can also be lowered through small eigenvalues, if
$\tilde{f}$ has a non-trivial matrix structure \cite{ji}.

\bibitem{grav} M. Y. Khlopov and A. D. Linde,  Phys. Lett. {\bf
B138} (1984) 265; J. R. Ellis, D. V. Nanopoulos and S. Sarkar,
Nucl. Phys. {\bf B259} (1985) 175; J. R. Ellis, D. V. Nanopoulos,
K. A. Olive and S. J. Rey, Astropart. Phys. {\bf 4}
(1996) 371; M. Kawasaki and T. Moroi, Prog. Theor. Phys.  
{\bf 93} (1995) 879; V. S. Rychkov and A. Strumia, hep-ph/0701104.

\bibitem{f3} The small logarithmic running between the electroweak scale and
the SUSY scale ($M_S$) and, in  effect, set $M_S$ and $M_Z$ to be
the same.

\bibitem{carena} M. Carena, S. Pokorski and C. E. M. Wagner, Nucl.
Phys. {\bf B406} (1993) 59.

\bibitem{baer} H. Baer, J. Ferrandis, S. Kraml and W. Porod, Phys.
Rev. {\bf D73} (2006) 015010.

\bibitem{langacker} P. Langacker and N. Polonsky, Phys. Rev. {\bf
D47} (1993) 4028.

\bibitem{parida1} M. K. Parida, B. Purkayastha, C. R. Das and B. D.
Cajee, Eur. Phys. J {\bf C28} (2002) 353; M. K. Parida and B. D.
Cajee, Eur. Phys. J {\bf C44} (2005) 447.

\bibitem{shafi} Q. Shafi and C. Wetterich, Phys. Rev. Lett. {\bf
52} (1984) 875; C. T. Hill, Phys. Lett. {\bf B135} (1984) 47; L.
Hall and U. Sarid, Phys. Rev. Lett. {\bf 70} (1993) 2673.

\bibitem{parpat} M. K. Parida and P. K. Patra, Phys. Rev. {\bf D39}
(1989) 2000; M. K. Parida and P. K. Patra, Phys. Lett. {\bf B432}
(1990) 45.

\bibitem{f4} Here, $M_S = M_Z$ has been assumed. 
If $M_S$ is set at 1 TeV, then one
finds $\R^0 = 5.0\times 10^{15} (1.6\times 10^{15})$  GeV and
$\U^0 = 1.9\times 10^{16} (6.2\times 10^{15})$ GeV, at the
one-loop  level in Model I (Model II).

\bibitem{ji} X. Ji, Y. Li, R. N. Mohapatra and S. Nasri, hep-ph/0605088; 
J. C. Pati, Phys Rev. {\bf D68} (2003) 072002.

\bibitem{albr} S. M. Barr, ref.{\cite {min4}}, C. H. Albright and S. M. Barr,
Phys. Rev. {\bf D69} (2004) 101601; K. L. McDonald, B. H. J. McKellar and A.
Mastrano, Phys.Rev. {\bf D70} (2004) 053012; D. G. Lee and R. N. Mohapatra, ref.{\cite{lee}}.

\bibitem{weinberg} S. Weinberg, Phys. Lett. {\bf B91} (1980) 51.

\bibitem{hall} L. Hall, Nucl. Phys. {\bf B178} (1981) 75; M.
Shifman, Int. J. Mod. Phys. {\bf A11} (1996) 5761.

\bibitem{ovrut} B. Ovrut and H. Schnitzer, {\bf B179} (1981) 381; W.
J. Marciano, in {\em Field Theory in Elementary Particles}, ed. A.
Perlmutter (Plenum, New York 1982), Proceedings of Orbis
Scientiae Vol. 19; in {\em Fifth Workshop on Grand Unification.
Philadelphia, Pennsylvania, 1983}, eds. H. A. Weldon, P. Langacker
and P. J. Steinhardt (Birkhauser, Boston, 1983); R. N. Mohapatra
and M. K. Parida, Phys. Rev. {\bf D47} (1993) 264.

\bibitem{f5} One must also ensure 
that the ratios ${M_i\over \U}, i=1,2,3$ lie within an appropriate
range, say 0.1 to 10, and ought not exceed the Planck mass.

\bibitem{felipe} D. Costa, P. Perez and R. Felipe, arXiv: hep-ph/0610178.


\bibitem{lee}D. G. Lee and 
R. N. Mohapatra, Phys. Rev. {\bf D52} (1995) 4125;
 M. Bando, J. Sato and T. Takahashi, Phys. Rev. {\bf D52} (1995) 3076;
E. Ma, Phys. Rev. {\bf D51} (1995) 236; E. Ma, Phys. Lett. {\bf B344}
(1995) 164.

\bibitem{babu} K. S. Babu and S. M. Barr, Phys. Rev. {\bf D48} (1993) 5354;
B. Dutta, Y. Mimura and R. N. Mohapatra, Phys. Rev. Lett. {\bf
94} (2005) 091804; Phys. Rev. {\bf D72} (2005) 075009.

\bibitem{fmsen} B. Bajc, A. Melfo, G. Senjanovi\'c and F. Vissani, 
Phys. Rev. {\bf D73} (2006) 055001; 
Phys. Lett. {\bf B634} (2006) 272.

\bibitem{f6} These states represent pseudo-Goldstone
bosons and may also acquire masses near the $\U$ scale through
loops.

\bibitem{f7} Here it has been assumed that the light scalar
components in $ \{15,3,1\} \oplus\{15,1,3\} \subset {\bf 210}$,
emerging from mass spectra predictions, are made superheavy. 
 This is possible if, for example, the minimal models are extended 
by the addition of a Higgs representation $\bf {54}$ in each case. But the
situation would be worse still in both the models if the scalar components  
remain light in the absence of  ${\bf 54}$ or suitable nonrenormalizable terms
in the superpotential.

\end{thebibliography}
\end{document}